\documentclass[12pt]{article}
\usepackage[dvips]{color}
\usepackage{epsfig}
\usepackage{amsmath}
\usepackage{graphicx}

\begin{document}
\title{Thermodynamics of string black hole with hyperscaling violation}
\author{{J. Sadeghi$^{a,b}$ \thanks{Email: pouriya@ipm.ir}\hspace{1mm} B. Pourhassan$^{a}$ \thanks{Email: b.pourhassan@umz.ac.ir}\hspace{1mm}
and A. Asadi$^{b}$ \thanks{Email: ali.asadi89@stu.umz.ac.ir}}\\
$^{a}${\small {\em Young researchers club, Ayatollah Amoli branch,
Islamic azad university, Amol, Iran}}\\
$^{b}${\small {\em Department of Physics, Mazandaran University,
Babolsar, Iran}}} \maketitle
\begin{abstract}
\noindent\\
In this paper, we start with black brane and construct specific
space-time which violates hyperscaling. In order to obtain the
string solution we apply Null-Melvin-Twist and $KK$-reduction. By
using the difference action method we study thermodynamics of system
to obtain Hawking-Page phase transition. In order to have
hyperscaling violation we need to consider $\theta=\frac{d}{2}.$  In
that case the free energy $F$ is always negative and our solution is
thermal radiation without a black hole. Therefore we find that there
is not any Hawking-Page transition. Also, we discuss the stability
of system and all
thermodynamical quantities.
\end{abstract}
\section{Introduction}
As we know the AdS/CFT correspondence provides an analytic approach
to study strongly coupled field theory
\cite{Maldacena1,Gubster,Witten,Aharony}. Recently,  we see several
paper about development of AdS gravity theories and their conformal
field theory dual, in that case the metric background generalized
and result is dual to scale-invariant field theories instead of
conformal invariant. The scale invariance provided by dynamical
critical exponent $z\neq1$ (the $z=1$ corresponds to case of the AdS
metric) on the following metric, \cite{Dong},
\begin{equation}\label{metric01}
ds^2=-\frac{1}{r^{2z}}dt^2+\frac{1}{r^2}\left(dr^2+dx^2_i\right),
\end{equation}
The corresponding metric will be invariant under following scale
transformation,
\begin{equation}\label{scale1}
t\rightarrow\lambda^zt,\qquad x_i\rightarrow\lambda x_i,\qquad
 r\rightarrow\lambda r.
\end{equation}
The resulting metric may be a solution of field equations with
coupled theories to matter with negative cosmological constant also
include an abelian field in the bulk. Space-time metrics that
transform covariantly under dilatation have recently been
reinterpreted as holography dual to stress tensor of quantum field
theories which violates hyperscaling \cite{Kachru,Tylor}. Recently,
the large class of scaling metrics containing an abelian gauge field
and scalar dilaton considered [7-19], which is presented by the
following equation \cite{Charmousis},
\begin{equation}\label{metric02}
ds^2=r^{-2{{(d-\theta)}/{d}}}\left(r^{-2(z-1)}dt^2+dr^2+dx^2_i\right),
\end{equation}
where $\theta$ is hyperscaling violation exponent. Note that, this
metric is not invariant under scale transformation (\ref{scale1}),
but transforms covariantly as,
\begin{equation}\label{scale2}
ds=\lambda^{\theta/d}ds,
\end{equation}
which defines property of hyperscaling violation in holography
language.  The corrections of conformal hyperscaling relation in the
conformal point of view in large $N_f$ QCD as a concrete dynamical
model is given by the Ref. \cite{Aoki}. Such examples show that QCD
can be a candidate for usage hyperscaling. On the other hand we have
strong motivation to study a metric with hyperscaling violation. As
we know, scale invariance broken under quantum effects in some
theories such as QCD. Specially in large scale or low energy, scale
invariance broken for massive theories. In that cases, using AdS
metric which has scale invariance is not appropriate. For example,
we have some problems to calculate form factor or quantum mass
spectrum in QCD. Therefore, it is necessary to modify original
metric [21]. Instead modification of AdS metric, it is appropriate
to choose suitable metric such as hyperscale metric which has scale
violation. In large scale (or $r\rightarrow\infty$) there are good
applications of a metric with hyperscaling violation in QCD or
string theory. So, in this paper we use metric of the Ref. [7] to
obtain string solution. Then, we discuss physical properties
(specially thermodynamics) of mentioned metric
to verify our motivation. If our solution will be coincide with known physical rules, then one can use hyperscale metric instead of AdS metric for future works.\\
An important concept in our study is the Galilean holography which
is developed in Refs. \cite{Son,Balasubramanian}, where the
non-relativistic generalizations of the AdS/CFT correspondence
extracted. An expansion of Galilean algebra can be obtained by
adding dilation operator and a special conformal transformation to
the time and space scale identically. A discussion of
non-relativistic conformal symmetry generalization which is known as
Schr\"{o}dinger has been explained in Ref. \cite{Witten}. In this
discussion, time and space geometry of $d$ dimensions isometry group
has Schr\"{o}dinger symmetry and established over AdS/CFT
correspondence. They suggested that the gravity is the holographic
dual of the non-relativistic conformal field theories at strong
couplings. The next development of Galilean holographic is finite
temperature generalization \cite{Maldacena2,Herzog,Adams}. In the
AdS/CFT correspondence of finite temperature a planar black brane
solutions suggested in the Schr\"{o}dinger space as the holographic
dual of the non-relativistic conformal field theory at finite
temperature. The investigation of $AdS_5$ geometry near horizon of
D3-brane in flat space is investigated \cite{Maldacena2,Herzog}.
Then, the known Null-Melvin-Twist (NMT)
\cite{Alishahiha,Gimon,Sadeghi} applied to this system.  Ref.
\cite{Adams} started with solution of asymptotical black hole
metrics which leads to the string solution and characterize the
specific non-relativistic conformal field theories to which they are
dual. An analysis of these black hole space-time thermodynamics
shows that they describe the dual conformal field theory at finite
temperature and finite density. It has been shown that, after doing
NMT by applying $KK$-reduction over $S^5$ geometry, the result is
extremal black brane also the asymptotic limits is reduced to
Schr\"{o}dinger geometry. The thermodynamic solutions of such black hole discussed by Refs. \cite{Maldacena2,Herzog}.\\
The new regularization method has been suggested by the Ref. \cite{Yamada} which is the oldest regularization method \cite{Hawking,Gibbons} with some
modification which is subtraction method with an unusual boundary matching.\\
In the recent work [33], thermodynamics of Schrödinger black holes
with hyperscaling violation has been studied. It may be found some
overlaps between our work and mentioned paper, however we should
note that our system is completely different and application of
hyperscaling violation in both systems yields to independent results
which are interesting itself. While the primary metrics of both
papers are similar but we use NMT method to obtain string solution,
also discuss about phase transition and thermodynamics stability.\\
This paper is organized as the following. In next section we begin
with black brane metric and make the corresponding metric which
violates hyperscaling. In that case, we apply NMT and $KK$-reduction
to obtain the string solution of this geometry. In section 3 we use
the difference action method and extract the thermodynamics of
system in section 4 and discuss Hawking-Page phase transition and
thermodynamics stability. In section 5 we summarized our results.
\section{String Black Brane}
Now, we consider the non-extremal D3-brane geometry \cite{Yamada}
near horizon, which is obtained by the following action
\cite{Son,Herzog},
\begin{eqnarray}\label{metric03}
&ds^2=\left(\frac{r}{R}\right)^2\left(-fdt^2+dy^2+dx^2_i\right)+\left(\frac{R}{r}\right)^2f^{-1}dr^2
+R^2d\Omega^2_5,\nonumber\\
\nonumber\\
&\phi=0,\qquad B=0,\qquad f(r)=1-\left(\frac{r_H}{r}\right)^4,
\end{eqnarray}
where $R$ is the AdS scale, $x_i=(x_1,x_2)$ and $r=r_H$ is the
location of the horizon, so the metric at $r_{H}=0$ reduces to the
extremal case. $\phi$ is dilaton and $B$ is $NS-NS$ two-form. A
particularly convenient choice for $d\phi$ is given by Hopf
fibration $s^1\rightarrow\ s^5\rightarrow\ p^2$ with the following
metric,
\begin{equation}\label{spher}
d\Omega^2_5 =ds^2_{p^2}+(d\chi+\mathcal{A}),
\end{equation}
where $\chi$ is the local coordinate on Hopf fibre and $\mathcal{A}$
is the one-form on $P^2$, and $ds^2_{p^2}$ is metric on $P^2$
\cite{Adams}. We need to consider two isometry directions as $dy$
and $d\phi$ for Melvinization process, where $dy$ is along the
world-volume, $d\phi$ is along the $S^5$ and $y$ is one of three
spatial coordinates. Now, we begin with the metric (5) includes
hyperscaling violation in the black hole solution according to the
Ref. \cite{Dong},
\begin{eqnarray}\label{metric04}
&ds^2_{d+2}=\left(\frac{r}{R}\right)^2\left(\frac{r_F}{r}\right)^{2\theta/d}\left(-\left(\frac{R^2}{r}\right)^{-2(z-1)}fdt^2+dy^2+dx^2_i\right)
+\left(\frac{R}{r}\right)^2\left(\frac{r_F}{r}\right)^{2\theta/d}\left(\frac{dr^2}{f}+R^2d\Omega^2_5\right),\nonumber\\
\nonumber\\
&f=1-\left(\frac{r_H}{r}\right)^{d+z-\theta},
\end{eqnarray}
where $d=3$, and $r_F$ is scale which is obtained from dimensional
analysis \cite{Dong}. Finite temperature effects in theories with
hyperscaling violation studied, in that case, in the gravity side,
we have $r_F<r_h$. From null energy condition $(NEC)$ as $T_{\mu\nu}
n^{\mu}n^{\nu}\geq 0$ \cite{Dong,Narayan} and null vectors satisfy
the $n^{\mu}n^{\nu}=0$ condition. The above conditions lead us to
obtain the following relations,
\begin{eqnarray}\label{NEC1}
&(d-\theta)(d(z-1)-\theta)\geq0,\nonumber\\
&(z-1)(d+z-\theta)\geq0.
\end{eqnarray}
In order to satisfy our following results with equation (\ref{scale2}) we need to consider $z=1$ (because $z=1$ in $\theta\rightarrow0$ limit gives the AdS
metric). From the first relation of (\ref{NEC1}), one can obtain,
\begin{equation}\label{NEC2}
(\theta\leq0\,\,\,,\qquad d\geq\theta),\qquad or \qquad
(\theta\geq0\,\,\,,\qquad d\leq\theta).
\end{equation}
Now, we apply NMT to the metric (\ref{metric04}) with $z=1$, and
obtain,
\begin{eqnarray}\label{metric05}
ds^2_{d+2} &=&
K^{-1}\left(\frac{r}{R}\right)^2M\left[-(1+b^2r^2M^2)fdt^2-2b^2r^2fM^2dtdy+(1-b^2r^2fM^2)dy^2+Kdx^2_i\right],\nonumber\\
&+&
M\left(\frac{R}{r}\right)^2f^{-1}dr^2+MK^{-1}R^2\eta^2+MR^2ds^2_{p^2},
\end{eqnarray}
and,
\begin{eqnarray}\label{field01}
&&\phi=-\frac{1}{2}\ln K,\nonumber\\
&&B=\frac{M^{2}}{K}\left(\frac{r}{R}\right)^2b(fdt+dy)\wedge\eta,\nonumber\\
&&K=1-(f-1)br^2M^2,
\end{eqnarray}
where $\eta=(d\chi+\mathcal{A})$,
$M=\left(\frac{r_F}{r}\right)^{(2\theta/d)}$, and also $b$ has
$[L^{-1}]$ dimension. If we perform the $KK$-reduction on $S^5$ for
the non-extremal solution (\ref{metric05}), we obtain,
\begin{eqnarray}\label{metric06}
ds^2_{d+2} &=& K^{-2/3}\left(\frac{r}{R}\right)^2M\left[-(1+b^2r^2M^2)fdt^2-2b^2r^2fM^2dtdy+(1-b^2r^2fM^2)dy^2+Kdx^2_i\right]\nonumber\\
&+& K^{1/3}M\left(\frac{R}{r}\right)^2f^{-1}dr^2,
\end{eqnarray}
and,
\begin{eqnarray}\label{field02}
&&\phi=-\frac{1}{2}\ln K,\nonumber\\
&&A=\frac{M^{2}}{K}\left(\frac{r}{R}\right)^2b(fdt+dy),
\end{eqnarray}
where $A$ is one-form field in Einstein frame. It is useful to work
in the following light-cone coordinates,
\begin{eqnarray}\label{light-cone}
x^{+}=bR(t+y),\qquad and,\qquad x^{-}=\frac{1}{2bR}(t-y).
\end{eqnarray}
So, the solution is,
\begin{eqnarray}\label{metric07}
ds^2_{d+2} &=& K^{-2/3}\left(\frac{r}{R}\right)^2M\bigg[-\left(\frac{f-1}{(2bR)^2}-\left(\frac{r}{R}\right)^2fM^2\right){dx^{+}}^2-(1+f)dx^{+}dx^{-}\nonumber\\
&+&
(bR)^2(1-f){dx^{-}}^2+Kdx^2_i\bigg]+K^{1/3}M\left(\frac{R}{r}\right)^2f^{-1}dr^2,
\end{eqnarray}
and,
\begin{eqnarray}\label{field03}
&&\phi=-\frac{1}{2}\ln K,\nonumber\\
&&A=\frac{M^{2}}{K}\left(\frac{r}{R}\right)^2b\left[\frac{f+1}{2bR}{dx^{+}}+bR(1-f)dx^{-}\right].
\end{eqnarray}
The equation (\ref{metric07}) is the same as the equation
(\ref{metric03}) in Ref. \cite{Yamada} with additional
$M=\left(\frac{r_F}{r}\right)^{2\theta/d}.$ By consideration $x^+$
coordinate as the time, the recent metric under scale transformation
$x^{+}\rightarrow\lambda^zx^{+},\,\,x_i\rightarrow\lambda x_i,\,\,
r\rightarrow\lambda^{-1} r,\,\,x_{-}\rightarrow\lambda^{2-z} x_{-}$
and $d=2\theta$ transforms covariantly as the equation
(\ref{scale2}), and it is violates hyperscaling.\\
The extremal case coming from $f=1$, and non-extremal case
approaches this at asymptotically large $r$. The last metric on the
light-cone coordinates in the equation (\ref{light-cone}) gives
extremal case which is independent of the parameter $b$. So, $b$ is
unphysical and thus cannot give any physical quantity. One can
interpret this result in the zero-temperature limit \cite{Adams}.
The metric background (\ref{metric07}) is a solution of the
effective action. In non-extremal case, for the $\theta=0$ we have
the following action \cite{Yamada},
\begin{equation}\label{action01}
S_5=\frac{1}{16\pi G_5}\int
dx^{5}\sqrt{-g}\left[\mathcal{R}-\frac{4}{3}(\partial_\mu\phi)(\partial^{\mu}\phi)-
\frac{1}{4}R^2e^{-8\phi/3}F_{\mu\nu}F^{\mu\nu}-4A_\mu
A^{\mu}-\frac{V}{R^2}\right],
\end{equation}
where $G_5$, $g$ and $R$ are the 5 dimensional Newton constant, the
determinant of 5 dimensional metric and the scalar curvature
respectively. $F=dA$ is two-form field and the potential $V$ is
defined by the following expression,
\begin{equation}\label{potential}
V=4e^{2\phi/3}(e^{2\phi}-4).
\end{equation}
By setting $\phi=0$, the above action reduces to the extremal action
\cite{Son}. As we know, in case of $\theta\neq0$, the shape of
action (\ref{action01}) will conserve, but in this process the
potentials
$V$ and corresponding field $\phi$ will be changed. Because the $K$ will be changed by parameter $\theta$.\\
The ADM form of metric is,
\begin{eqnarray}\label{metric09}
ds^2_{d+2}&=&K^{1/3}\left(\frac{r_F}{r}\right)^{(2\theta/d)} \left(\frac{R}{r}\right)^2f^{-1}dr^2\nonumber\\
&+&K^{-2/3}\left(\frac{r}{R}\right)^2\left(\frac{r_F}{r}\right)^{(2\theta/d)}\bigg[Kdx^2_i-\left(\frac{1}{(bR)^2(1-f)}
+\left(\frac{r}{R}\right)^2\left(\frac{r_F}{r}\right)^{(4\theta/d)}\right) f{dx^{+}}^2\bigg]\nonumber\\
&+&K^{-2/3}\left(\frac{r}{R}\right)^2\left(\frac{r_F}{r}\right)^{(2\theta/d)}\bigg[(bR)^2(1-f)\left(dx^{-}-\frac{(1+f)}{2(bR)^2(1-f)}dx^{+}\right)^2\bigg].
\end{eqnarray}
By using the corresponding metric, we obtain the angular velocity of
the horizon $\Omega_H$, which interpreted as chemical potential
associated with the conserved quantities along the $x^{-}$
direction,
\begin{equation}\label{a.v}
\Omega_H=\frac{1}{2(bR)^2}.
\end{equation}
Note that we have mentioned two kinds of hypersurfaces; the
time-like boundary at a large fixed $r$ and the space-like surface
at a fixed time $x^{+}$ whose time is described by the ADM form. In
the  extremal case there is a problem with $g_{--}$ component in
calculation of difference action $(g_{--}=0)$.
\section{The Difference Action}
The  metric (\ref{metric07}) gives the extremal solution near the
boundary (the large $r$) and  interpreted as the finite temperature
generalization of the Galilean holography
\cite{Maldacena2,Herzog,Adams}. We want to consider the
thermodynamics of this system in the finite temperature. In order to
calculate the thermodynamics,  we use difference action method
\cite{Yamada,Hawking,Gibbons}.\\
According to the Ref. \cite{Yamada}, first we continue analytically
$x^{+}$ to $ix^{+}$ and put the system into a box by cutoff $r=r_B$.
The cutoff $r_B$ is larger than the scale $R$ but it is finite. We
subtract the action of the extremal solution from the non-extremal
one. We note here each action include two terms such as bulk and
Gibbons-Hawking surface term. To do such process, we have to match
the geometries of metrics in $r=r_B$ wall. As mentioned earlier, the
$g_{--}$ component of the extremal case has been degenerated in the
metric (\ref{metric07}), so we cannot match metrics in the wall. In
order to remove this problem, we match the boundary metric of the
extremal geometry to the non-extremal one only for the $x^{-}$
constant. So, we rescale appropriately three dimensional slices
$(x^{+},x^{i})$. We obtain scaled extremal metric as a following,
\begin{eqnarray}\label{metric10}
ds^2_{d+2} &=&
\left(\frac{r}{R}\right)^2\left(\frac{r_F}{r}\right)^{(2\theta/d)}
\left[\left(\frac{r}{R}\right)^2\left(\frac{r_F}{r}\right)^{(4\theta/d)}H^2_B{dx^{+}}^2-2iH_Bdx^{+}dx^{-}
+G^2_B{dx_{i}}^2\right]\nonumber\\
&+& \left(\frac{R}{r}\right)^2\left(\frac{r_F}{r}\right)^{(2\theta/d)}dr^2,\nonumber\\
\phi &=& 0,\nonumber\\
A &=& i\left(\frac{r}{R}\right)^2\left(\frac{r_{F}}{r}\right)^{\frac{4\theta}{d}}\frac{H_B}{R}dx^{+},
\end{eqnarray}
where,
\begin{eqnarray}\label{func}
H_B &=& \left[K(r_B)^{-2/3}\left(\frac{f(r_B)-1}{(2bR)^2}+
{\left(\frac{r_B}{R}\right)}^2\left(\frac{r_F}{r_B}\right)^{(4\theta/d)}f(r_B)\right)\right]^{1/2}
\left(\frac{r_B}{R}\right)^{-1}\left(\frac{r_F}{r_B}\right)^{(-2\theta/d)},\nonumber\\
G_B &=& K\left(r_B\right)^{1/6}.
\end{eqnarray}
The difference action $(S-S_0)$ will be as,
\begin{eqnarray}\label{action02}
S_{0}=S_{0bulk}+S_{0GH},\qquad and ,\qquad S=S_{bulk}+S_{GH},
\end{eqnarray}
where both $S_{0bulk}$ and $S_{bulk}$ are action (\ref{action01}),
but the $S_{0bulk}$ evaluate on the extremal solution
(\ref{metric10}) and the $S_{bulk}$ calculate on the non-extremal
solution (\ref{metric07}). Also  $S_{0GH}$ and $S_{GH}$ are the
Gibbons-Hawking surface term,
\begin{equation}\label{action03}
S_{0GH}=-\frac{1}{8\pi G_5}\int dx^{4}\sqrt{g_B}\left(TrK_0\right),
\end{equation}
where $g_B$ is the determinant of the boundary first fundamental
form, and $(TrK_0)$ is the trace of the boundary second fundamental
form. We calculate the difference action in the limit of
$r_B\rightarrow\infty$, which is not divergent,
\begin{equation}\label{d.action01}
\lim_{r_B\rightarrow\infty}(S-S_0)=\frac{V_4}{16\pi
G_5}\frac{r^4_H}{R^5}\left(1-\frac{\theta}{d}\right)\left(\frac{r_F}{r_H}\right)^{3\theta/d},
\end{equation}
where $V_4$ is volume of four dimensions space-time. It shown that this result agree with Ref. \cite{Yamada} without hyperscaling violation.
\section{Thermodynamics}
Now, we use results of the previous section to study the thermodynamics of system. In that case the Hawking temperature can be obtained from surface
gravity as $\beta=\frac{2\pi}{\kappa}$ where $\kappa$ is surface gravity,
\begin{equation}\label{s.gr}
\kappa^2=-\frac{1}{2}\left(\nabla^{a}\xi^{b}\right)\left(\nabla_{a}\xi_{b}\right),
\end{equation}
where $\xi$ is the killing vector field which is obtained by following expression,
\begin{equation}\label{killing}
\xi=\frac{1}{bR}\frac{\partial}{\partial
t}=\partial_{+}+\Omega_H\partial_{-},
\end{equation}
and corresponding $\beta$ is obtained by,
\begin{equation}\label{beta}
\beta=\frac{4}{d+1-\theta}\frac{\pi bR^3}{r_H}.
\end{equation}
The killing generator of the event horizon (\ref{killing}) not only
has components along the boundary time translation direction
$x^{+}$, but also along light-like direction $x^{-}$. From the
gravitational point of view it is therefore a system with chemical
potential for $x^{-}$ directions,
\begin{equation}\label{chemical}
\mu=\frac{1}{2(bR)^2}.
\end{equation}
In order to study the thermodynamics of system,  we use the
following free energy \cite{Yamada,Rostami,Pourdarvish},
\begin{eqnarray}\label{f.e}
F &=& -(16\pi{G_5})V^{-1}_3\lim_{r_B\rightarrow\infty}(S-S_0)\nonumber\\
&=& -\beta\left(\frac{r^4_H}{R^5}\right)\left(1-\frac{\theta}{d}\right)\left(\frac{r_F}{r_H}\right)^{3\theta/d} \nonumber\\
&=&
-\frac{\pi^4R^3}{4\mu^2\beta^3}\left(1-\frac{\theta}{d}\right)\left(\frac{\beta
r_F}{\pi
R^2}\right)^{3\theta/d}\left(\frac{4}{d+1-\theta}\right)^{4-3\theta/d}(2\mu)^{3\theta/2d},
\end{eqnarray}
where $V_3$ is the integration over $x^{-,i}$, and equal to $V_4
\beta^{-1}$. So we obtain entropy as,
\begin{eqnarray}\label{entropy}
S &=& \beta\left(\frac{\partial F}{\partial\beta}\right)_\mu-F\nonumber\\
&=& \frac{4\pi
br^3_H}{R^2}\frac{(1-\frac{\theta}{d})}{(d+1-\theta)}\left(4-\frac{3\theta}{d}
\right)\left(\frac{r_F}{r_H}\right)^{3\theta/d}.
\end{eqnarray}
These equations, in the case of $\theta=0$, agree with the Ref.
\cite{Yamada}. Also we can obtain,
\begin{eqnarray}\label{energy}
E &=& \left(\frac{\partial F}{\partial
\beta}\right)_\mu-\mu\beta^{-1}\left(\frac{\partial F}{\partial
\mu}\right)_\beta \nonumber\\
&=& \frac{r^4_H}{R^5}\left(1-\frac{\theta}{d}\right)\left(1-\frac{3\theta}{2d}\right)\left(\frac{r_F}{r_H}\right)^{3\theta/d},\nonumber\\
Q &=& -\beta^{-1}\left(\frac{\partial F}{\partial\mu}\right)_\beta\nonumber\\
&=&
-\frac{4b^2r^4_H}{R^3}\left(1-\frac{\theta}{d}\right)\left(1-\frac{3\theta}{4d}\right)\left(\frac{r_F}{r_H}\right)^{3\theta/d}.
\end{eqnarray}
In equation (\ref{f.e}) we have two conditions for $F$ such $F>0$
and $F<0$. In the case of $F<0$ we have two conditions as $\theta<d$
or $\theta>d+1$, and in the case of $F>0$ we have $d<\theta<d+1$.
So, in the case of $\theta=d$ and $\theta=d+1$ we have Hawking-Page
phase transition. As mentioned before we take $\theta=d/2$, so we
have always negative $F$. So, our solution is thermal radiation
without a black hole and we have not any Hawking-Page phase
transition. As we know, in order to calculate the stability of
system we need to obtain the Hessian of $\beta (E-\mu Q)-S$ with
respect to the thermodynamic variables $(r_H, b)$ and evaluate it at
the on-shell values of $(\beta, \mu)$. In the case of $\theta=0$ it
recovers the results of the Ref. \cite{Yamada}. In the case of
hyperscaling violation with condition of $\theta>\frac{1}{2}$, the
results will be positive and the system is thermodynamically stable.
Here, also we check the first law as $dE=TdS +\Omega_H {dQ}$ and
satisfy by the above quantities.
\section{Conclusion}
In this paper, we considered the black brane metric and made
corresponding metric which is violate hyperscaling. By using the
difference action method we obtained the thermodynamical quantities
such as $\beta, Q, S, E$, and $F$. In the case of $F>0$ we achieved
two conditions as $\theta<d$ or $\theta>d+1.$ And also for $F>0$ we
arrived at $d<\theta<d+1$. Two above conditions lead to Hawking-Page
phase transition $(\theta=d, \theta=d+1)$. But in this paper we have
always negative $F$ because our condition was $\theta=\frac{d}{2}$
and we have not such phase transition. Also we discussed the
stability of system which agree with the Ref. \cite{Yamada} in
$\theta=0$. We have shown that in the case of hyperscaling violation
the $\theta$ must be $\theta>\frac{1}{2}$ which is covered by our
condition. In general we can say that the system has thermodynamical
stability. Therefore one can use hyperscale metric instead of AdS
metric to avoid technical problems in boundary because of scale
symmetry breaking. For future work we focus on this subject and use
a hyperscale metric instead of AdS metric to calculate form factor
in QCD. Finally we verified that the first law of thermodynamics is
valid.

\end{document}